# Chalcogenide Perovskites – an Emerging Class of Ionic Semiconductors

Samanthe Perera[a,1], Haolei Hui[a,b,1], Chuan Zhao[a], Hongtao Xue[a], Fan Sun[a], Chenhua Deng[c], Nelson Gross[a], Chris Milleville[d], Xiaohong Xu[c], David F. Watson[d], Bernard Weinstein[a], Yi-Yang Sun[e,*], Shengbai Zhang[e] and Hao Zeng[a,*]

[a]Department of Physics, University at Buffalo, the State University of New York, Buffalo NY 14260, USA

[b]School of Science, Xi'an Jiaotong University, Xi'an, 710049, China

[c]Key Laboratory of Magnetic Molecules and Magnetic Information Material of Ministry of Education, School of Chemistry and Materials Science of Shanxi Normal University, Linfen 041004, China

[d]Department of Chemistry, University at Buffalo, the State University of New York, Buffalo, NY 14260, USA

[e]Department of Physics, Applied Physics & Astronomy, Rensselaer Polytechnic Institute, Troy, New York 12180, USA

[1]These authors contributed equally to this work.
**\*** Corresponding Authors: suny4@rpi.edu, haozeng@buffalo.edu

**ABSTRACT**

We report the synthesis and characterization of a novel class of ionic semiconductor materials – inorganic chalcogenide perovskites. Several different compounds including $BaZrS_3$, $CaZrS_3$, $SrTiS_3$ and $SrZrS_3$ were synthesized by high temperature sulfurization of their oxide counterparts. Their crystal structures were identified by XRD and composition by EDX. UV-vis and photoluminescence measurements confirmed that they are direct gap semiconductors with band gap values consistent with theoretical predictions. By adopting an anion alloying approach, we demonstrate widely tunable band gap from 1.73 eV to 2.87 eV. These strongly ionic semiconductors provide a new avenue for engineering the semiconducting properties for applications such as energy harvesting, solid state lighting and sensing.

Semiconductor materials have wide ranging industrial applications from microelectronics to energy harvesting and sensing. Conventional semiconductors that dominate the current industry are mainly covalent materials, as characterized by four-fold coordination of all atoms. Recently, the photovoltaic research community has focused their attention on an unconventional material: the organic-inorganic halide perovskite. The power conversion efficiency (PCE) of perovskite solar cells has experienced phenomenal increase from an initial value of 3.8% in 2009 [1] to 15% in 2013 [2-5], and then quickly to 21.0% in another two years [6-8][9]. Perovskites, however, are ionic materials as characterized by higher coordination maximizing the Coulomb interaction between cations and anions. It is likely that this ionicity could be a major contributing factor for the excellent transport properties of the halide perovskites, such as the extremely long carrier diffusion length [10-12]. It has been reported that in such materials deep level defects are less likely to form [13, 14]. The great success of the halide perovskite materials has inspired us to search for novel semiconductor materials that can retain the excellent optoelectronic properties of halides, while avoiding their severe limitations such as instability and toxicity [15, 16].

As opposed to their oxide and halide siblings, chalcogenide perovskites have received little attention, despite being synthesized more than half a century ago [17, 18]. Publications on these materials are surprisingly scarce [19-21], and as a result, there is limited knowledge of their physical properties [22, 23]. Our recent theoretical study [24] have screened 18 $ABX_3$ chalcogenide materials (A=Ca, Sr, or Ba; B=Ti, Zr, or Hf; X=S or Se) with several exciting discoveries: 1) These materials in a distorted perovskite structure are semiconductors with band gaps ranging from 0.3 to 2.3 eV; 2) They could possess direct band gaps with high optical absorption coefficients, comparable to those of traditional optoelectronic semiconductors such as GaAs; 3) The band dispersion is large for those Zr and Hf compounds, suggesting good carrier mobility. With these predicted properties, these materials are particularly attractive for photovoltaic and optoelectronic applications and therefore call for experimental verification.

In this paper, we report the synthesis of polycrystalline chalcogenide perovskites (BaZrS$_3$, CaZrS$_3$, SrTiS$_3$ and SrZrS$_3$) and the characterization of their structural and physical properties. These materials were synthesized by sulfurization of oxide perovskites by CS$_2$ at high temperatures. X-ray diffraction (XRD) shows that BaZrS$_3$ possesses a distorted perovskite structure. Low temperature Raman measurements reveal multiple modes ranging from 50 to 400 cm$^{-1}$, in good agreement with theoretical predictions. UV-vis absorption spectroscopy confirms that it is a semiconductor, with high absorbance from UV to visible region. The band gap value is estimated to be 1.73 eV, again in good agreement with theoretical predictions. The sample also show photoluminescence (PL) in the visible region, confirming its direct band gap. This material shows excellent stability in ambient conditions as compared to halide perovskites. We further demonstrate systematic band gap engineering by anion alloying in oxysulfides.

**Experimental**

**Synthesis of BaZrS$_3$.** A ceramic boat containing 0.6 g of BaZrO$_3$ (≥99% purity, Alfa Aesar) was heated at a rate of 15 °C/min in a 2" quartz tube furnace under flowing Ar. CS$_2$ (≥99.8% purity, Alfa Aesar) was introduced at 800 °C by bubbling Ar through a gas bubbler containing CS$_2$. Once the temperature reached 1050 °C, it was kept constant for 4 hours before cooling down to 800 °C at the same rate. At 800 °C, CS$_2$ flow was turned off and the temperature was kept constant for another 2 hours under Ar flow to remove unreacted sulfur, before cooling down to room temperature. The exhaust gas from the tube was bubbled through 1M NaOH solution throughout the reaction to remove excess CS$_2$. The CS$_2$ flow rate was adjusted from 10 to 20 standard cubic centimeters per minute (sccm). The reaction temperature and time were optimized to obtain phase pure $BaZrS_3$ without oxides or other impurity phases.

**Synthesis of oxysulfides.** Seven $BaZrO_3$ powder samples were placed in the uniform temperature region in a two zone furnace at an interval of 1 cm apart inside the quartz tube. The sample closest to upstream is labeled as S1 and the one farthest is labeled as S7. The synthesis was carried out similar to that of $BaZrS_3$. The flow rate of $CS_2$ was kept at 15 sccm. By limiting the $CS_2$ flow and controlling the position of the $BaZrO_3$ powder, sulfur deficiency was enforced in these samples. $BaZrO_3$ located the closest to the upstream (S1) was completely sulfurized, while S7 remains nearly un-sulfurized; in between, oxysulfides $BaZr(O_xS_{1-x})_3$ with various degree of sulfurization were achieved.

**Crystal structure and phase information** was acquired using a Rigaku Ultima IV X-ray diffraction (XRD) system operating at 1.76 kW (40 kV, 44mA) with Cu target for X-ray source and dual position graphite diffracted beam monochromator.

**UV-vis absorption spectra** were obtained by measuring diffuse reflectance with a wavelength ranged from 400 nm to 1100 nm using Labsphere RSA-HP-8453 reflectance spectroscopy accessory attached to Agilent 8453 ultra-violet/visible (UV-vis) spectroscopy system.

**Raman and PL measurements** Stokes Raman was collected using RCA photomutiplier tube attached to a Jobin Yvonne U1000 double monochromator with 1800 lines/mm holographic gratings. For excitation 4-5 mw of the 676.457 nm line from a Coherent Innova 100 $Kr^+$ ion laser was focused with a 300 mm lens to produce a spot size of ~130 μm. Spectra were averaged three times and smoothed with 5 point adjacent averaging.

**Morphology and energy-dispersive X-ray elemental analysis** was carried out using a Hitachi SU70 field emission scanning electron microscope (FESEM) with Oxford energy-dispersive X-ray spectrometer (EDS) running a silicon drift detector.

**First-principles calculation** was based on density functional theory using the PBEsol functional [25] to optimize the lattice constants and atomic coordinates. Band structure was calculated using the HSE06 hybrid functional [26]. Raman spectrum calculations (frequency and intensity) were carried out following Ref. [27] using the PBEsol functional.

## Results and discussions

Typical SEM images of the polycrystalline $BaZrO_3$ precursor and $BaZrS_3$ powder sample are shown in Fig. 1(a) and 1(b), respectively. The $BaZrO_3$ precursor has a grain size of 100-200 nm. After sulfurization, the particles become much larger of a few μm in size. The particles are faceted suggesting good crystallinity. The color of $BaZrO_3$ is white (inset of Fig. 1(a)), suggesting no optical absorption at the visible range, consistent with its band gap value of 5 eV. After sulfurization, the sample becomes black (inset of Fig. 1(b)), suggesting visible light absorption in a broad range. EDX analysis (Fig. 1(c)) shows a composition close to stoichiometry, with the atomic ratios of Ba: Zr: S =21.6: 21.4: 57.0. The slight off-stoichiometry from $ABX_3$ is not uncommon in perovskites, and mainly reflects the presence of sulfur vacancies due to high temperature treatment. The XRD spectrum of $BaZrS_3$ is shown in Fig. 2. It can be seen that all peaks can be matched to those of the standard file JCPDS 00-015-0327. This confirms that the sample possesses an orthorhombic distorted perovskite structure with Pnma space group. The lattice constants are measured to be $a = 7.04$ Å, $b = 9.98$ Å and $c = 7.05$ Å, respectively, closely matched to the published results [18]. No secondary phase is detected within the instrument limit.

Raman spectra measured at different temperatures from 14 to 295 K are shown in Fig. 3(a). At room temperature, roughly 5 broad peaks can be observed. With decreasing temperature, these 5 peaks are progressively sharpened due to the customary decrease in anharmonic thermal phonon-decay. As a result, the broad peak centered at about 215 cm$^{-1}$ is shown to consist of two peaks at 216 and 224 cm$^{-1}$, which are identified to be $A_g^6$ and $B_{2g}^6$ modes, respectively. At the reduced measuring temperatures of 14-150K, the sharpening concentrates the mode oscillator strengths into enhanced peak heights permitting weaker additional features to be detected, *e.g.,* at 169 cm$^{-1}$ and 304 cm$^{-1}$. The low temperature Raman measurements allow direct comparison with first-principles calculations. According to group theory, orthorhombic Pnma structure should have 24 Raman modes, $7A_g + 5B_{1g} + 7B_{2g} + 5B_{3g}$ [28-30]. The calculated peak position and intensity are shown in Fig. 3(b), together with the experimental spectrum measured at 14 K. In total, 14 Raman modes can be identified with theoretical assignments given in Table 1. The peak positions show good agreement with experimental data. On the other hand, as seen from Fig. 3(c), BaZrO$_3$ shows no Raman peak at this frequency range, which is expected since BaZrO$_3$ is cubic (space group Pm3m) with no significant first-order Raman scattering [31]. The Raman results, combined with XRD results, show that our BaZrS$_3$ is phase pure with good crystalline quality.

**Table 1.** Assignment of experimental Raman peaks according to the irreducible representations of point group $D_{2h}$ based on calculated frequencies. The unit is cm$^{-1}$.

| mode | Theory | Exp. | mode | Theory | Exp. | mode | Theory | Exp. |
|---|---|---|---|---|---|---|---|---|
| $B_{3g}^5$ | 418.1 | -- | $B_{3g}^3$ | 203.0 | 207 | $A_g^3$ | 98.8 | 97 |
| $B_{2g}^7$ | 399.1 | -- | $B_{2g}^5$ | 201.9 | 207 | $B_{2g}^3$ | 97.6 | 93 |
| $B_{1g}^5$ | 388.0 | 392 | $B_{1g}^3$ | 174.5 | -- | $B_{1g}^1$ | 85.8 | 83 |
| $A_g^7$ | 304.6 | 304 | $A_g^5$ | 168.7 | 169 | $B_{2g}^2$ | 81.3 | 78 |
| $B_{1g}^4$ | 290.8 | -- | $B_{2g}^4$ | 167.8 | 169 | $B_{3g}^1$ | 76.9 | 72 |
| $B_{3g}^4$ | 290.3 | -- | $B_{3g}^2$ | 160.2 | -- | $A_g^2$ | 76.0 | 72 |
| $B_{2g}^6$ | 214.8 | 224 | $A_g^4$ | 142.7 | 143 | $B_{2g}^1$ | 64.0 | 64 |
| $A_g^6$ | 207.0 | 216 | $B_{1g}^2$ | 111.0 | -- | $A_g^1$ | 58.7 | 57 |

Figure 4 shows the band structure of BaZrS$_3$ calculated using the HSE06 hybrid functional, which suggests that BaZrS$_3$ is a direct band gap semiconductor with both conduction band minimum and valence band maximum located at the Γ point. The band gap is 1.74 eV. It can also be seen that the bottom of conduction band has relatively large dispersion, hence good electron transport, even though it is mainly derived from the transition metal element Zr. To verify the electronic structure calculations, UV-vis spectroscopy and PL have been performed. As can be seen from the UV-vis absorption spectrum in Fig. 5, BaZrS$_3$ has an absorption tail extending to low energies, making accurate determination of the band gap difficult [32]. The band gap value is estimated to be about 1.73 eV, in good agreement with the theoretical prediction. As can be seen from the PL spectrum plotted in the same graph, a broad PL peak centered at about 1.7 eV can be observed. The emission near the band gap energy confirms that BaZrS$_3$ is a direct band gap material. The absorption and PL results suggest that BaZrS$_3$ has desirable optical properties for energy harvesting and other optoelectronic applications.

We further demonstrate band gap engineering by anion alloying. Since BaZrO$_3$ is an insulator with a band gap of about 5 eV while that of BaZrS$_3$ is about 1.8 eV. We postulate that the band gap can be systematically tuned by formation of oxysulfides with different O$^{2-}$ to S$^{2-}$ ratio. To achieve this, we intentionally tuned the chemical potential to favor the formation of oxysulfides, by limiting the flow of $CS_2$ to ensure sulfur deficiency, while keeping the temperature at 1,050 °C to allow thermodynamically stable phase to be formed. With such an approach, the composition of the chalcogenide perovskites can be symmetrically tuned from essentially oxides ($BaZrO_3$) to nearly 100% sulfides ($BaZrS_3$). As can be seen from the EDX analysis (Table 2), the composition of S and O shows opposite trends from S1 to S7: while S increases from 3.4 % for S1 to 55.8% for S7, O decreases from 51% to close to 0, respectively.

**Table 2.** The atomic percentage of the 4 elements (Ba, Zr, O and S) for samples S1 to S7, as measured by EDX spectroscopy.

|     | S1    | S2    | S3    | S4    | S5    | S6    | S7    |
| --- | ----- | ----- | ----- | ----- | ----- | ----- | ----- |
| **Ba** | 25.08 | 25.09 | 23.42 | 21.98 | 25.54 | 25.64 | 23.79 |
| **Zr** | 20.68 | 20.9  | 21.26 | 20.31 | 21.48 | 22.23 | 22.31 |
| **O**  | 50.85 | 46.55 | 37.1  | 33.68 | 23.59 | 19.74 | 0     |
| **S**  | 3.38  | 7.46  | 18.22 | 24.03 | 29.4  | 32.4  | 53.89 |

The color of the samples from S1 to S7 changes progressively from nearly white to completely black, as shown by the optical images of these samples in Fig. 6(a). UV-vis absorption spectra (Fig. 6(b)) show that with increasing S content, the optical absorption progressively red-shifts to longer wavelengths (lower energies). The estimated band gap values are plotted as a function of S concentration in Fig. 6(c). It is seen that $E_g$ decreases monotonically from 2.87 eV for S1 to 1.75 eV for S7. It is straightforward to understand the tuning of the band gap: since the valence band of these perovskites consists of mainly the p states of O/S anions, and O 3$p$ states are deeper in energy than that of O 2$p$ states, replacing $O^{2-}$ with $S^{2-}$ effectively narrows the band gap by uplifting the valence band minimum (VBM). Nevertheless, it is fascinating that the band gap can be systematically tuned over a wide range by the degree of sulfurization. Such range is of great interest to broad applications from energy harvesting, solid state lighting to sensing.

Other chalcogenide perovskites including $CaZrS_3$, $SrZrS_3$ and $SrTiS_3$ have also been synthesized using similar techniques. $CaZrS_3$ is found to be in distorted perovskite structure with a band gap of 1.90 eV (Fig. 6(b)), while both $SrTiS_3$ and $SrZrS_3$ possess a needle-like phase. Both are predicted to have very low band gap values beyond the range of our UV-vis spectrometer. The stability of $BaZrS_3$ was tested by measuring the XRD pattern and UV-vis absorption spectrum of a sample stored in air for 7 months. No detectable change of properties has been observed (Fig. S3). Repeated washing of $BaZrS_3$ by deionized water also does not result in change of its properties.

In conclusion, several chalcogenide and oxysulfide perovskite compounds including $BaZrS_3$, $CaZrS_3$, $SrZrS_3$, $SrTiS_3$ and $BaZr(O_xS_{1-x})_3$ have been synthesized. These materials belong to a new class of ionic semiconductors. The band gap of these materials can be systematically tuned in a wide range from UV to infrared. Due to their predicted strong iconicity they may exhibit unique physical properties such as free of deep level defects, which is beneficial for energy harvesting and other optoelectronic applications. It should be pointed out that oxide perovskites, with a chemical formula of $ABO_3$, has long been an active field of research. This family of materials exhibits unusually rich properties ranging from colossal magnetoresistance, ferroelectricity to superconductivity and charge density waves, resulting from interplay of different degrees of freedoms with similar energy scales. The intriguing physics of their chalcogenide counterparts, however, is largely unexplored. Developing chalcogenide perovskite materials therefore not only provide new materials complementing oxides, but also offers new opportunities for exploring new physical phenomena and material properties tunable by light, electric and magnetic fields.

**Acknowledgement:**

Work supported by NSF DMR-1104994, CBET-1510121, and MRI-1229208. HLH supported by the Education Program for Talented Students of Xi'an Jiaotong University.

**Figure captions:**

Figure 1. SEM images of the polycrystalline (a) BaZrO$_3$ powder used for the synthesis and (b) the synthesized BaZrS$_3$ powder sample; (c) the EDX spectrum of BaZrS$_3$. Insets of Fig. 1(a) and (b) are the optical images of the white BaZrO$_3$ powder and black BaZrS$_3$ powder, respectively. The scale bars are 5 μm.

Figure 2. The XRD pattern of the BaZrS$_3$. All peaks can be matched to the standard file JCPDS 00-015-0327 of the orthorhombic structured BaZrS$_3$ (space group Pnma).

Figure 3. (a) The Raman spectra of BaZrS$_3$ measured at different temperatures from 14 to 295 K; (b) the calculated Raman modes shown together with the spectrum measured at 14 K; (c) Raman spectrum of BaZrO$_3$.

Figure 4. Top view (a) and side view (b) of BaZrS$_3$ in the distorted perovskite structure (space group Pnma). Red dashed frames and black arrows mark the unit cell and the crystallographic orientations, respectively. (c) Band structure of BaZrS$_3$ calculated using the HSE06 hybrid functional. The high-symmetry *k*-points X, Y, Z, U, S, R, and Γ correspond to ($\pi/a$, 0, 0), (0, $\pi/b$, 0), (0, 0, $\pi/c$), ($\pi/a$, 0, $\pi/c$), ($\pi/a$, $\pi/b$, 0), ($\pi/a$, $\pi/b$, $\pi/c$), and (0, 0, 0), respectively, in the Brillouin zone. The rightmost panel shows the total density of states (DOS) and the site-projected DOS on Zr 4d and S 3p orbitals.

Figure 5. The UV-vis absorption spectrum and PL spectrum of BaZrS$_3$.

Figure 6. (a) The optical images and (b) the UV-vis absorption spectra of the BaZr-oxysulfides with different degrees of sulfurization; and (c) the measured band gap energy as a function of sulfur atomic concentration.

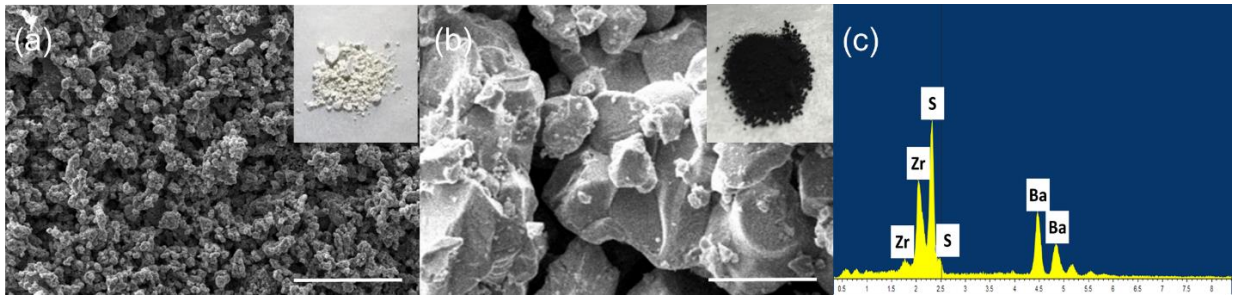

Figure 1

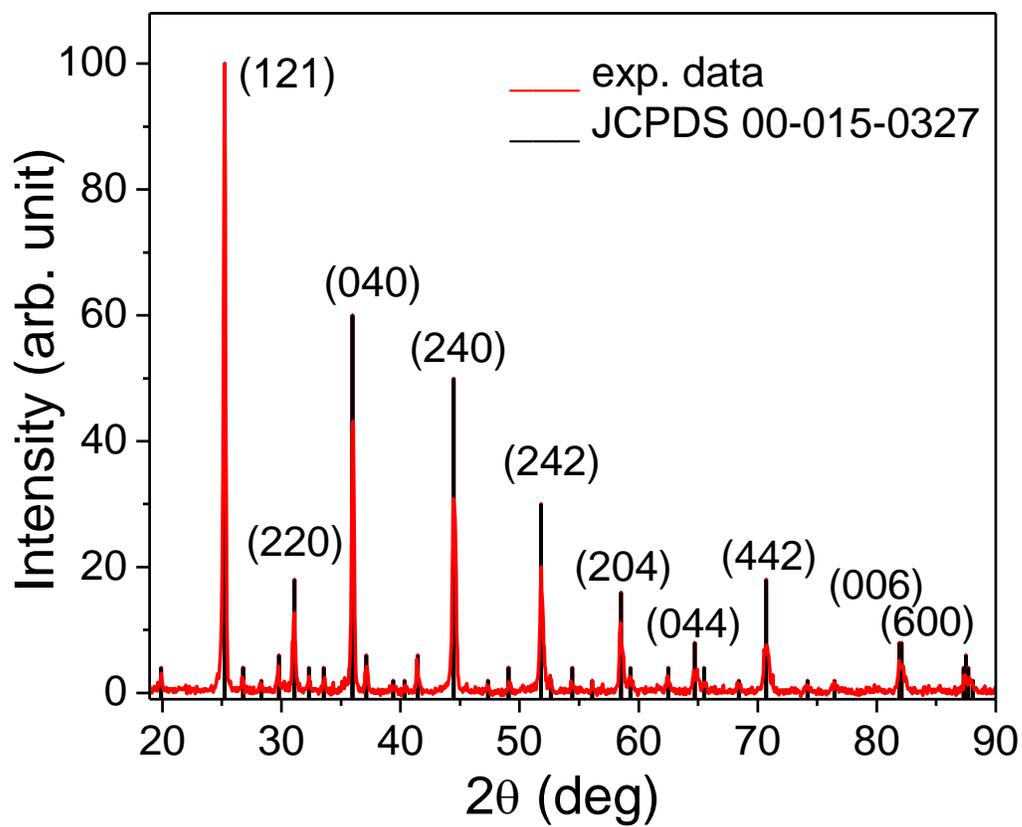

Figure 2

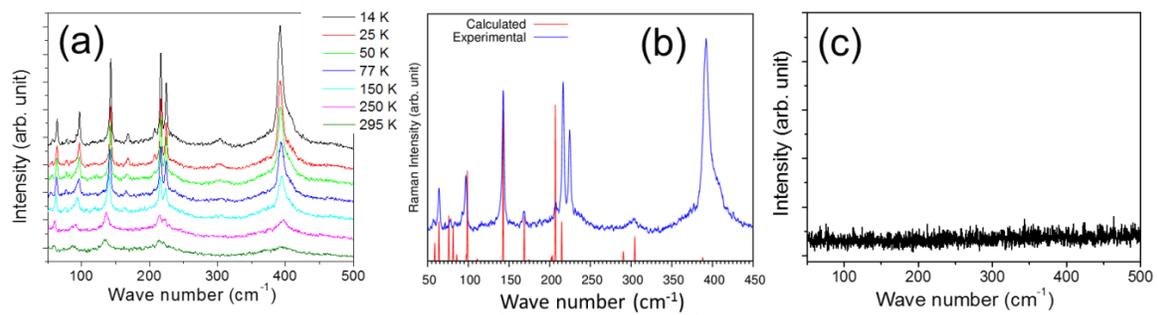

Figure 3

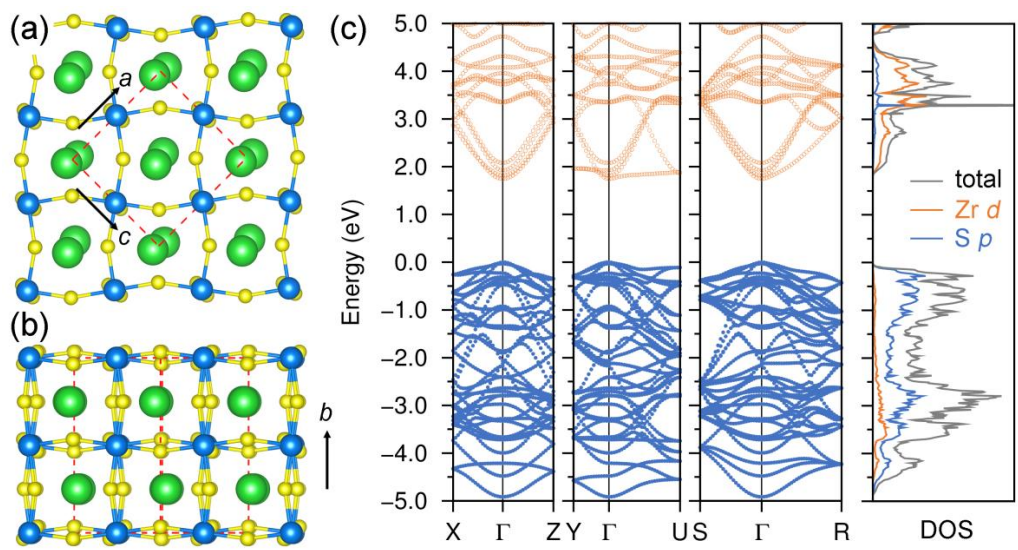

Figure 4

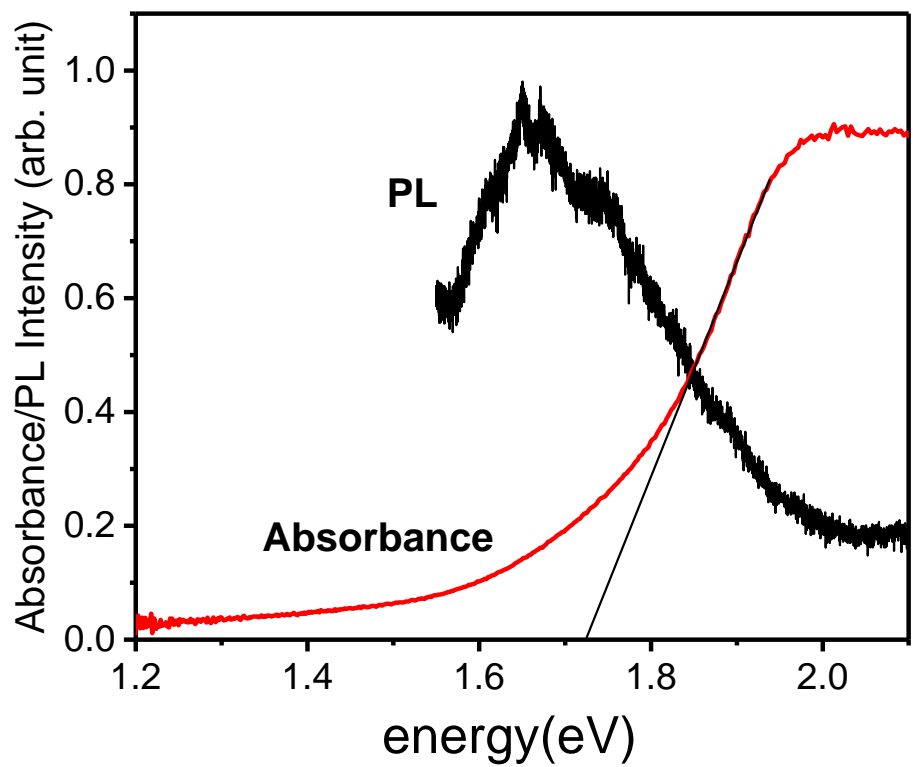

Figure 5

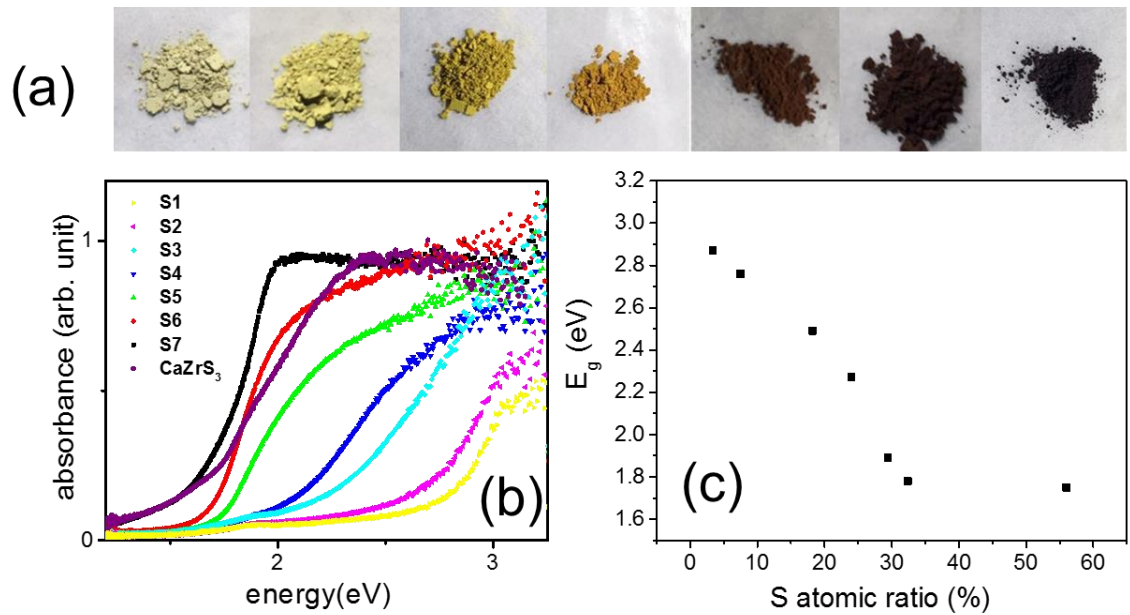

Figure 6